\documentstyle[sprocl,epsf]{article}

\def\Journal#1#2#3#4{{#1} {\bf #2}, #3 (#4)}

\def\NPA{{\em Nucl. Phys.} A}

\def\PLB{{\em Phys. Lett.}  B}
\def\PRL{\em Phys. Rev. Lett.}
\def\PRD{{\em Phys. Rev.} D}

\def\ZPA{{\em Z. Phys.} A}

\def\ANP{{\em Adv. in Nucl. and Part. Phys., eds. J.Negele and E. Vogt}}

\def\ubar{\bar{u}}
\def\dbar{\bar{d}}

\def\be{\begin{equation}}
\def\ee{\end{equation}}
\def\bea{\begin{eqnarray}}
\def\eea{\end{eqnarray}}

\begin{document}

\title{THE IMPACT OF NLO CORRECTIONS ON THE DETERMINATIONS OF THE $\bar{u},\bar{d}$ CONTENT
OF NUCLEONS FROM DRELL-YAN PRODUCTION}

\author{W. SCH\"AFER}

\address{Institut f\"ur Kernphysik, Forschungszentrum J\"ulich, D-52425 J\"ulich}


\maketitle\abstracts{The interpretation of Drell-Yan production in terms of the
  antiquark densities depends on NLO corrections. Besides the NLO corrections to the
  familiar annihilation $q\bar{q}\to \gamma^* \to l^+ l^-$, there is a substantial
  contribution from the QCD Compton subprocesses $gq \to q\gamma^* \to q l^+ l^-$ and
  $g\bar{q} \to q\gamma^* \to q l^+ l^-$. The beam and target dependence of the two
  classes of corrections is different. We discuss the impact of this difference on
  the determination of the $\bar{d}-\bar{u}$ asymmetry in the proton from the
  comparison of the $pp$ and $pn$ Drell-Yan production.}

The substantial $\bar{d}-\bar{u}$ asymmetry of the nucleon sea is a striking
manifestation of the leading twist nonperturbative mesonic effects in parton
distributions (for the recent review and references see \cite{Speth}). The first
indirect experimental evidence for this asymmetry has been deduced from the Gottfried
sum rule evaluation \cite{GSR}, but it left of course open the $x-$dependence of the
$\bar{d}-\bar{u}$ asymmetry. Following a suggestion by Ellis and Stirling\cite{Stir}
, the CERN NA51 collaboration has measured the $\bar{d}-\bar{u}$ asymmetry at $x \simeq 0.18$
from the comparison of the $pp$ and $pn(pd)$ Drell-Yan production. The asymmetry
proved to be large and placed at large $x$, in good agreement with expectations from
meson cloud models \cite{GSR,DY}.
Recently, a much more detailed and higher precision comparison of
the $pp$ and $pd$ Drell-Yan production has been performed in the Fermilab E866
experiment. Furthermore there are plans of the dedicated Drell-Yan experiment at
Fermilab with the goal of measuring the flavour content of the sea up to $x \sim
0.6$\cite{Garvey}.  The accuracy of the E866 experiment is so high that one must wonder about the
impact of next-to-leading order (NLO) corrections on the comparison of Drell-Yan
production on different targets.  The point is that one would like to interpret the
Drell-Yan process in terms of the annihilation of quarks from the beam hadron on
antiquarks in the target hadron.  Roughly speaking, in a suitable kinematic domain,
Drell-Yan production is dominated by annihilation of the valence $u$-quarks of the
beam proton on the $\bar{u}$-sea in the target proton. Because the valence densities
are well known, in the $pp$-Drell-Yan production one measures the $\ubar$ sea.Invoking the
charge symmetry, the $pn$-Drell-Yan measures the $\dbar$ sea and the ratio of the $pn$
and $pp$ cross sections gives an access to the $\dbar / \ubar$ ratio. NLO
corrections spoil this simple picture.  They fall into two broad categories
\cite{RvN}: the QCD radiative corrections to the familiar annihilation (ANN) $q
\bar{q} \to \gamma^* \to l^+ l^-$ retain the property of the Drell-Yan cross section
being a (sum over flavours of the) product of quark and antiquark densities. The
second class of corrections however comes from the 'QCD-Compton' (QCDC) subprocesses
$gq \to q\gamma^*$ and $g\bar{q} \to \bar{q} \gamma^*$ and has the form of a certain
convolution of the gluon and quark+antiquark densities in the beam and target and
vice versa.  The difference in the beam and target parton contents of the two classes
of corrections and their dependence on the form of the parton densities raise several
issues: Firstly it is well established that NLO corrections, ususally parametrised in
terms of the so-called $K_{DY}$-factor,$K_{DY} = d\sigma_{DY}(LO+NLO)/
d\sigma_{DY}(LO)$, are large: $K_{DY}\approx 1.5 - 2$. Second, the $x$-dependence of
the gluon and sea densities is different, and there is a troubling possibility that
the relative size of the contributions from annihilation and QCDC NLO corrections
changes over the space of the beam and target Bjorken variables. Furthermore the
NLO corrections depend on the shape of parton distributions, and may change the
realative importance of the $u \ubar$ and $d \dbar$ annihilation contributions
compared to the LO formulas.  To summarize, 
the issue is that NLO corrections are large, and because in the E866 data
the statistical accuracy of the measured $pn/pp$ ratio is at the level of several
percent, a scrutiny of those subtleties of the NLO corrections is called upon \cite{NSS}.

\begin{figure}[h]
\epsfxsize=1.0\hsize
\epsfbox{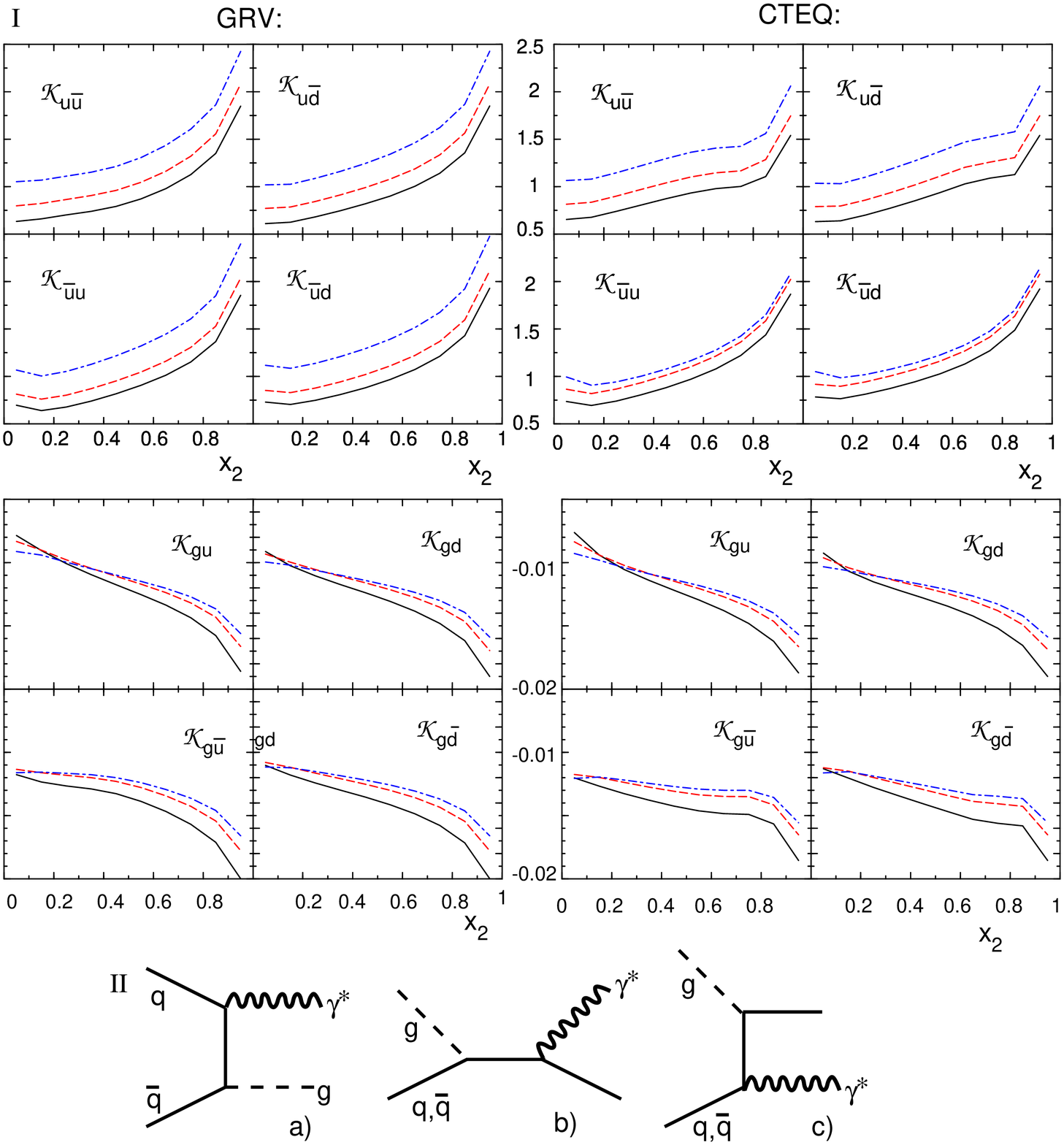}
\caption{I: A selection of partial ${\cal{K}}$--factors calculated with two representative 
parametrizations of parton densities; shown as function of $x_2$ for several $x_1$--values: $x_1 = 0.35$ (solid); $x_1 = 0.55$ (dashed); $x_1 = 0.75$ (dot-dashed);i.e. $x_1$ grows from bottom to top curves. II:  a) gluon bremsstrahlung correction to the $q\bar{q}$ annihilation, b) QCDC contributions.}
 \label{fig1}
\end{figure}

The purpose of the present study is an investigation of the beam and target
dependence of NLO corrections relevant to the interpretation of the $pn/pp$ data; we
shall be primarily interested in the region of rather large $x$ relevant to the
E866-data.  The contributions from the NLO corrections to the differential cross
section appear in the form \bea {d\Delta \sigma_{A} \over d Q^{2}dx_{F} } \propto
\sum_{f} e_{f}^{2} \Delta_{f\bar{f}}(t_1,t_{2},x_{1},x_{2},Q^{2})\otimes
\left[f_{1}(t_{1},Q^{2})\bar{f}_{2}(t_{2},Q^{2}) +(1 \leftrightarrow 2)\right]
\label{eq:NLOannih}
\eea
for the annihilation corrections of Fig.~1 IIa); and similarly for the QCDC corrections
(fig 1 IIb,c):
\bea
{d\Delta\sigma_{C} \over d Q^{2}dx_{F} } \propto
\left\{
\Delta_{fg}(t_{1},t_{2},x_{1},x_{2},Q^{2}) \otimes
\left (\sum e_{f}^{2}\left[ f_{1}(t_{1},Q^{2})+ \bar{f}_{1}(t_{1},Q^{2})
\right]\right)
g_{2}(t_{2},Q^{2})\right.\nonumber\\
+
\left. \Delta_{gf}(t_{1},t_{2},x_{1},x_{2},Q^{2}) \otimes g_{1}(t_{1},Q^{2})
\left(\sum e_{f}^{2}\left[f_{2}(t_{2},Q^{2}) +
\bar{f}_{2}(t_{2},Q^{2})\right]\right)\right\} \, .\nonumber
\eea
\label{eq:NLOQCDC}
The convolution sign here implies an integration in the variables $t_i, i=1,2$ over
the domain $x_i < t_i < 1$ and all the required coefficient functions
$\Delta_{f\bar{f}}, \Delta_{fg}$
are available in the
literature \cite{RvN}.  For our purposes it is most convenient to study the
dependende on flavour etc.  in terms of partial ${\cal{K}}$--factors, that
parametrise the NLO-corrections in the following manner: \bea
\Delta_{f\bar{f}}(t_1,t_{2},x_{1},x_{2},Q^{2}) \otimes
f_{1}(t_{1},Q^{2})\bar{f}_{2}(t_{2},Q^{2})  = \nonumber \\{\cal
  K}_{f\bar{f}}(x_{1},n_{f},x_{2},n_{\bar{f}}) \cdot
f_{1}(x_1,Q^{2})
\bar{f}_{2}(x_{2},Q^{2})\nonumber \\ \Delta_{fg}(t_1,t_{2},x_{1},x_{2},Q^{2}) \otimes
f_{1}(t_{1},Q^{2})g_{2}(t_{2},Q^{2}) = \nonumber \\{\cal
  K}_{fg}(x_{1},n_{f},x_{2},n_{g})\cdot
f_{1}(x_{1} ,Q^{2}) (1-x_{2})g_{2}(x_{2},Q^{2})\,
.\nonumber \eea The functions ${\cal K}_{f\bar{f}},{\cal K}_{fg}$ now bear a
dependence on the \emph{functional form} of the parton distributions which is
encoded here through the exponents $n_f, n_g$ of the
large-$x$ behaviour, $f(x,Q^2) \sim (1-x)^{n_f}, g(x,Q^2) \sim (1-x)^{n_g}$. 
The above shown expressions enter the
calculation of the experimentally measured cross section, and the $n_f,
n_g$--dependence clearly demonstrates the potential model dependence of the
extraction of $\ubar, \dbar$-densities from experimental data. Let us emphasize that
there is no theoretical a priori reason that the QCDC contributions should
mimick the $q \bar q$--annihilation processes over a wide $(x_1,x_2,Q^2)$--range.
In figure \ref{fig1} we show a variety of partial $\cal{K}$--factors for two 
representative parametrizations of the parton distributions. A substantial dependence 
on flavour, $x_1$ and $x_2$ can clearly be seen. As a matter of fact, for 
$x_{1,2} > 0.5-0.6$ the parton distributions are so poorly known, that the calculations in that region have little reliability. In particular the interpretation of Drell-Yan experiments at larger 
$x_{1,2}$ values shall be strongly model dependent.
In the region of $x_{1,2}$ relevant to the E866 experiment however the model dependence 
turns out to be weak within the experimental error bars and an extraction of $\dbar/\ubar$ in
the LO formalism can still be trusted.



\end{document}